\documentclass[twocolumn,amssymb,amsmath,floats,showpacs,superscriptaddress,pre,floatfix]{revtex4-1}
\usepackage{amsmath,amssymb,bm}
\usepackage{graphics}
\usepackage{epsfig}
\usepackage{graphicx}

\begin{document}

\title{
Critical behavior of the relaxation rate, the susceptibility, and a pair correlation function in the Kuramoto model on scale-free networks}
% Force line breaks with \\

\author{S. Yoon}
\affiliation{Departamento de F\'isica $\&$ I3N, Universidade de Aveiro, Aveiro, Portugal}
\author{M. Sorbaro Sindaci}
\affiliation{Dipartimento di Fisica, Universit\`a di Pavia, Pavia, Italy}
\affiliation{Institute for Adaptive and Neural Computation, School of Informatics, University of Edinburgh, UK}
\author{A. V. Goltsev}
\affiliation{Departamento de F\'isica $\&$ I3N, Universidade de Aveiro, Aveiro, Portugal}
\affiliation{A. F. Ioffe Physico-Technical Institute, 194021 St. Petersburg, Russia}
%\email{goltsev@ua.pt}
\author{J. F. F. Mendes}
\affiliation{Departamento de F\'isica $\&$ I3N, Universidade de Aveiro, Aveiro, Portugal}

%\date{\today}

\begin{abstract}
We study the impact of network heterogeneity on relaxation dynamics of the Kuramoto model on uncorrelated complex networks with scale-free degree distributions.
Using the Ott-Antonsen method and the annealed-network approach, we find that the critical behavior of the relaxation rate near the synchronization phase transition does not depend on network heterogeneity and critical slowing down takes place  at the critical point when the second moment of the degree distribution is finite.
In the case of a complete graph we obtain an explicit result for the relaxation rate when the distribution of natural frequencies is Lorentzian. We also find a response of the Kuramoto model to an external field and show that the susceptibility of the model is inversely proportional to the relaxation rate. We reveal that network heterogeneity strongly impacts a field dependence of the relaxation rate and the susceptibility when the network has a divergent fourth moment of degree distribution. We introduce a pair correlation function of phase oscillators and show that it has a sharp peak at the critical point, signaling emergence of long-range correlations.
Our numerical simulations of the Kuramoto model
%on scale-free networks
support our analytical results.
\end{abstract}


%05.10.-a, 05.40.-a, 05.50.+q, 87.18.Sn
\maketitle

%\tableofcontents

\section{Introduction}
\label{intr}

%From periodic motions of planets to those of electrons around atoms, to mechanical vibrations, we are surrounded by oscillating systems. The importance of oscillation is not limited to physics: this is perhaps one of the reasons why physicists and mathematicians have grown interest towards multidisciplinary
%investigations of oscillations in many fields. As far as oscillatory systems hold such an
%importance in the mathematical description of reality, it is natural to consider interactions in the systems.
%We are here considering a particular kind of interaction, namely an interaction
%affecting \emph{synchrony} and, in particular, its \emph{spontaneous emergence}.
%%

Spontaneous emergence of synchronization is a well-known phenomenon which has been observed in a wide variety of real systems: the natural spectacle of fireflies flashing at the same time or crickets chirping in unison; circadian rhythms; opinion formation in social science, and then again in biology, physics, sociology and neuroscience \cite{Pikovski2001,StrogatzDivulg,Acebron2005,Arenas2008,breakspear2010}.
The model introduced by Kuramoto \cite{Kuramoto1975} is the most studied statistical model of  spontaneous synchronization.
It describes the evolution of the system of phase oscillators towards synchrony or disorder by means of a set of time differential equations,
%%%
\begin{equation}
\frac{d \theta_i}{dt} = \omega_i + \frac{K}{N}\sum_{j=1}^N \sin(\theta_j-\theta_i),
\label{k1}
\end{equation}
where $\theta_i$ and $\omega_i$ are the phase and  the `natural' frequency of an oscillator $i$, $i=1,2,\dots N$. $K$ is the coupling.
%(see a review \cite{Acebron2005}).
The values of each $\omega_i$ are usually extracted from a given probability density function $g(\omega)$. The important property of the model  is that, despite the heterogeneity in natural frequencies, the oscillators become spontaneously synchronized if $K$ is larger than a critical value. The synchronization order parameter is
\begin{equation}
z = re^{i \psi} \equiv \frac{1}{N } \sum_{j=1}^N e^{i\theta_j},
\label{op-o}
\end{equation}
where $\psi$ is the order parameter phase. The absolute value $r=|z|$ represents the degree of synchronicity, being equal to 0 when oscillators phases are uniformly distributed in $[0,2 \pi)$, and 1 when they all have the same phase.

The set of Eqs. (\ref{k1}) describes oscillators on a complete graph.
A far more realistic model is to set the system on a complex network, as it was recently done \cite{Lee2005,Ichinomiya2004,Dorogovtsev2008,Arenas2008}. In this model, apart the heterogeneity in natural frequencies, there is a structural heterogeneity that is a natural attribute of real systems. Understanding how this kind of heterogeneity impacts synchronization dynamics of oscillators is still elusive. The framework of networks allows to use well-known methods of the complex network theory that has been shown to be applicable to a striking number of real systems in science and engineering \cite{Newman2003,Dorogovtsev2008,Dorogovtsev2010}.
Unfortunately, it is difficult to describe a collective process of synchronization or desynchronization of large number of oscillators in the Kuramoto model by solving the set of $N$ non-linear equations (\ref{k1}).
However, using the method proposed by Ott and Antonsen \cite{Ott2008,Ott2009}, one can reduce the original set of differential equations to a differential equation describing temporal behavior of the order parameter Eq.~(\ref{op-o}) alone, eliminating all angular variables.
This methods allows to directly study relaxation dynamics of the order parameter and to answer the question: \emph{How long does the system fall in the equilibrium, to synchronize or de-synchronize}?
It was already shown that the relaxation time of the Kuramoto model
becomes infinite when the system approaches a critical point of a second-order synchronization phase transition (see for example \cite{strogatz1991stability,Daido_1990,Ott2008}), a phenomenon known as critical slowing down.
However, analytically, the relaxation rate of the Kuramoto model was only found in disordered state in the case of all-to-all coupling \cite{strogatz1991stability,Ott2008}. The impact of network heterogeneity on relaxation dynamics of the Kuramoto model on complex network in both disordered and synchronized state was not yet studied analytically.
This problem is nontrivial since it has already been shown that critical behavior of the order parameter is strongly influenced by network heterogeneity when the Kuramoto model has a complex network structure with fat tailed degree distributions
\cite{Lee2005,Ichinomiya2004,Dorogovtsev2008}.
Investigations of relaxation dynamics of the Kuramoto model is also interesting because the relaxation rate is related to susceptibility of the system to external stimulations.
In statistical physics there are general relationships between the relaxation rate, the susceptibility, and the pair correlation function characterizing correlations between distant interacting agents in a system \cite{Stanley_book}.
Within the mean-field theory, near a critical point of a second order phase transition, the zero-field susceptibility of a system is inversely proportional to the relaxation rate. Therefore, critical slowing down simultaneously  signals the  divergence of the susceptibility at the critical point. Moreover, critical behavior of the susceptibility can shows us is the synchronization transition mean-field like in the presence of network heterogeneity or not? In turn, the susceptibility is related to a pair correlation function and divergence of the function manifests simultaneously the emergence of long-rang correlations in the system. However, these relationships were not discussed in the context of the Kuramoto model in complex networks.

In this paper, using the Ott-Antonsen method and the annealed-network approach \cite{Dorogovtsev2008,Bianconi2002}, we study relaxation dynamics of the Kuramoto model
on uncorrelated complex networks with a scale-free degree distribution $P(q)\propto q^{-\gamma}$. At first,  we find the relaxation rate below and above the critical coupling and demonstrate critical slowing down at the critical point in networks with $\gamma > 3$.
In the case of all-to-all interaction and the Lorentz distribution of natural frequencies, we obtain an explicit dependence of the relaxation rate on the coupling. Then we study the Kuramoto model in an external field and find critical behavior of the susceptibility.
Our approach allows us to analyze a field dependence of the relaxation rate and the susceptibility at the critical point. We show that these parameters are described by power laws with power law exponents dependent on $\gamma$ when $3 < \gamma \leq 5$.
Finally, we introduce a pair correlation function of phase oscillators and study its critical behavior. In order to  support our analytical results, we also perform numerical simulations of the Kuramoto model on scale-free networks and compare with the analytical results.

\section{Annealed network approximation}
\label{ANA}

The Kuramoto model on a complex network is described by an equation,
\begin{equation}
\frac{d \theta_i}{dt} = \omega_i + K\sum_{j=1}^N a_{ij}\sin(\theta_j-\theta_i),
\label{a1}
\end{equation}
that is a natural generalization of Eq.~(\ref{k1}). Here $a_{ij}$ are the entries of the network's adjacency matrix:
\begin{equation}
a_{ij} = \begin{cases} 1 & \text{if $i$ is connected to $j$,}\\ 0 & \text{otherwise}. \end{cases}
\label{a2}
\end{equation}
However, Eq. (\ref{a1}) is too complex to obtain significant analytical results.
A good approximation is given by substituting the values $a_{ij}$ with their expected values $\langle a_{ij} \rangle\in[0,1]$ in a given ensemble of graphs \cite{Dorogovtsev2008,Bianconi2002}. For the case of a sparse uncorrelated complex network with a degree distribution $P(q)$,
\begin{equation}
\langle a_{ij} \rangle = \frac{q_{i} q_{j}}{N \langle q \rangle}
\label{a3}
\end{equation}
holds in the continuum ($N \to \infty$) limit, so that
\begin{equation}
\frac{d \theta_i}{dt} = \omega_i + \frac{K q_i}{N \langle q \rangle}\sum_{j=1}^N q_{j} \sin(\theta_j-\theta_i).
\label{kur_net}
\end{equation}
This so-called `annealed network' approximation gives correct critical behavior of phase transitions in complex networks  \cite{Dorogovtsev2008,Bianconi2002}, including the synchronization transition in the Kuramoto model \cite{Dorogovtsev2008,Coutinho2012}.
Within this approach \cite{Dorogovtsev2008,Bianconi2002}, a complex order parameter is defined as follows:
\begin{equation}
z = re^{i \psi} \equiv \frac{1}{N \langle q\rangle} \sum_{j=1}^N q_{j}e^{i\theta_j}.
\label{op1}
\end{equation}
%In this case
Here, the contribution of single oscillators is weighed on the degree of the node they are associated to. This choice, which is reasonable as it accounts for the role of the oscillators in the network, also allows to write (\ref{kur_net}) as
\begin{equation}
\frac{d \theta_i}{dt} = \omega_i - K q_{i} r \sin(\theta_i-\psi).
\label{kur_net_simp}
\end{equation}

\section{Dynamical approach}
\label{dynamics}

To obtain an explicit set of differential equations for the time evolution of $z$, we use
%a dimensional reduction method introduced by
the Ott-Antonsen method \cite{Ott2008,Ott2009}. First, in the thermodynamical limit,
%one can work with
we introduce the oscillator density $F(\theta,\omega,q,t)$ on a $(\theta,t)$ configuration space, with dependence on the parameters $\omega$ and, in the network case, $q$. By definition, it satisfies the following normalization conditions:
\begin{equation}
\int_{1}^\infty \int_{0}^{2 \pi} F(\theta,\omega,q,t) d \theta d q = g(\omega),
\label{d1}
\end{equation}
and
\begin{equation}
\int_{-\infty}^{+\infty}\int_{0}^{2 \pi} F(\theta,\omega,q,t) d \theta d \omega = P(q).
\label{d2}
\end{equation}
The fundamental equation is simply the conservation law of the number of oscillators:
\begin{equation}
\frac{\partial}{\partial t}F(\theta,\omega,q,t) + \frac{\partial}{\partial \theta}[v F(\theta,\omega,q,t)] = 0,
\label{d3}
\end{equation}
where $v$ is the velocity field on the circle that drives the dynamics of $F$. This is specific of the model, and in our case can be read from equation (\ref{kur_net_simp}) as
\begin{equation}
v(\theta,\omega,q,t) = \omega + K q r \sin(\psi-\theta) = \omega + \frac{K q}{2 i}(z e^{-i \theta}- z^*e^{i \theta}).
\label{d4}
\end{equation}

Following Ott-Antonsen method, a solution is sought in the form
\begin{equation}
F = \frac{P(q)g(\omega)}{2 \pi}( 1 + F_{+} +  F_{-} )
\label{d5}
\end{equation}
with
\begin{equation}
F_+ \equiv \sum_{n=1}^\infty F_n(\omega,q,t)e^{i n \theta}
\label{d6}
\end{equation}
and $F_{-}=F_{+}^*$. With the additional ansatz
\begin{equation}
F_n = \alpha^n(\omega,q,t),
\label{d7}
\end{equation}
one obtains an equation for the function $\alpha(\omega,q,t)$,
\begin{equation}
\dot{\alpha} + i \omega \alpha + \frac{K q}{2}(z \alpha^2- z^*) = 0
\label{alpha}
\end{equation}
which is not in a closed form yet. It can be shown that Eq.~(\ref{op1}) takes a form
\begin{equation}
z(t) = \int \frac{q P(q)}{\langle q \rangle} \int g(\omega) \alpha^*(\omega,q,t) d \omega d q.
\label{d8}
\end{equation}
These two equations (\ref{alpha}) and (\ref{d8}) form a closed system of equations for the complex order parameter $z$.

Let us study a steady state solution before studying the relaxation dynamics.
By stationary, here, we mean $z(t) = re^{i\psi+i\Omega t}$, for the constant order parameter $r$, a phase $\psi$, and a group angular velocity $\Omega$. With a suitable change of the reference frame, $\omega \mapsto \omega + \Omega$ and putting $\psi=0$, without loss of generality we have $z=r$ and stationary points of Eq. (\ref{alpha}) can be found by simply putting $\dot{\alpha}=0$ in Eq.~(\ref{alpha}). We find a solution
\begin{equation}
\alpha_0(\omega,q) = \begin{cases} -\frac{i \omega}{Kqr} +\sqrt{1-\left(\frac{\omega}{Kqr}\right)^2}, &  |\omega|\leq Kqr \\
-\frac{i\omega}{Kqr}\Bigl[1 - \sqrt{ 1-\left(\frac{Kqr}{\omega}\right)^2} \,\,\Bigr] & \text{otherwise}.
\end{cases}
\label{d9}
\end{equation}
This solution satisfies the requirement $|\alpha|\leq 1$ necessary to the convergence of the geometrical series Eq.~(\ref{d7}). Note that substitution of Eq.~(\ref{d9}) into Eq.~(\ref{d7}) gives explicitly the oscillator density Eq.~(\ref{d5}).
Real and imaginary parts of Eq.~(\ref{d8}) give, respectively,
\begin{eqnarray}
r &=& \int \frac{d q  P(q) q}{\langle q\rangle} \int_{-Kqr}^{+Kqr} d\omega g(\omega{+}\Omega)\sqrt{1{-}\left(\frac{\omega}{Kqr}\right)^2},
\label{eq:scrq} \\
\Omega &=& \langle\omega\rangle  {-}\! \int \!\!  \frac{d q P(q) q}{\langle q\rangle} \!\!\! \int_{|\omega|\geq Kqr} \!\!\!\!\!\!\!\!\!\!\! d\omega \operatorname{sgn}(\omega) g(\omega {+} \Omega) \sqrt{\omega^2{-}(Kqr)^2} .
\nonumber \\
\label{eq:omq}
\end{eqnarray}
%The equation (\ref{eq:omq}) determines the group frequency $\Omega$.
The integral over $\omega$ in Eq. (\ref{eq:omq}) equals zero if the frequency distribution $g(\omega)$ is symmetrical around its mean. In this case, symmetry reasons impose $\Omega = \langle \omega \rangle$. However, the model is more general, and extends to non-trivial networks the results of Basnarkov and Urumov \cite{Basnarkov2008}, who studied the case of asymmetric frequency distributions. An analysis of Eq. (\ref{eq:scrq}) and the critical behavior of the order parameter $r$ in complex networks with different topologies are represented in Appendix \ref{app:orderparameter-K}.

\section{Relaxation dynamics}
\label{rexation}
Once one has found the equilibrium points for the system --- and therefore the critical point --- what other information can be extracted from the equations? A natural question for the model
%in which a stable state is expected
is: How long does it take to attain synchrony (or disorder)? We aim to answer this question by studying the dynamical behavior of the system weakly perturbed from a stationary state Eq.~(\ref{d9}) with the order parameter $z_0=r$:
\begin{eqnarray}
z(t) &=& z_0 + \delta z(t),  \nonumber \\
\alpha(t) &=& \alpha_0(\omega,q) + \delta\alpha(\omega,q,t).
\label{d11}
\end{eqnarray}
In an initial state at $t=0$,  a parameter $\delta z(t=0)$ is determined by a function $\delta\alpha(\omega,q,t=0)$.  Excluding second and higher order terms in $\delta\alpha$ and $\delta z$ in Eq.~(\ref{alpha}), we obtain a linear equation for $\delta\alpha(\omega,q,t)$,
\begin{equation}
\delta\dot{\alpha} +  i \omega \delta \alpha + \frac{Kq}{2}(2z_0\alpha_0 \delta \alpha + \alpha_0^2 \delta z - \delta z^*) = 0,
\label{d12}
\end{equation}
that must be solved self-consistently with an equation,
\begin{equation}
\delta z(t) = \int \frac{q P(q)}{\langle q \rangle} \int g(\omega) \delta \alpha^*(\omega,q,t) d \omega d q.
\label{dz}
\end{equation}
To solve it, we take the Laplace transform of both sides and integrate by parts. Reordering of the terms then yields
\begin{equation}
\delta\alpha (s,\omega,q) = \frac{\delta \alpha(t =0) + \frac{Kq}{2}(\delta z^*(s)-\alpha_0^2 \delta z(s))}{s +  i \omega + Kq z_0 \alpha_0}.
\label{d13}
\end{equation}
Substituting $\delta\alpha (s,\omega,q)$ into Eq.~(\ref{dz}) leads to
\begin{equation}
\delta z(s) = \frac{B(s)}{1-\frac{K}{\langle q \rangle}\int dq  P(q) q^2 A(s,q)},
\label{d15}
\end{equation}
where
\begin{eqnarray}
\!\!\!\!\!\! B(s) &{=}& \frac{1}{\langle q \rangle} \!\! \int \!\! dq P(q) q^2 \int\!\! d\omega \frac{g(\omega{+}\Omega)\delta \alpha(\omega,q,t{=}0)}{s {+}  i \omega {+} K q r \alpha_0},
\label{d14aa} \\
\!\!\!\!\!\!A(s,q) &{=}& \frac{1}{2}\int_{-\infty}^{\infty} d\omega \frac{g(\omega+\Omega)[1-\alpha_0^2(\omega,q)]}{s +  i \omega + K q r\alpha_0(\omega,q)}.
\label{As1}
\end{eqnarray}
%%%%
%Here, $\alpha_0(\omega,q)$ and $r$ are given by Eqs.~(\ref{d9}) and (\ref{eq:scrq}).
To invert the Laplace transform Eq.~(\ref{d15}), we first have to find the poles of $\delta z(s)$, equating the denominator to zero:
\begin{equation}
\frac{1}{\langle q \rangle}\int dq  P(q) q^2 A(s,q) = \frac{1}{K}.
\label{d17}
\end{equation}
It is important to note that the function $A(s,q)$ must be calculated at a positive $s$ and then analytically continued to negative values.

\subsection{Relaxation rate below the critical coupling, $K< K_c$}
\label{relaxationbelow}

In the region below the critical point, $K < K_c$,
the Kuramoto model is in a disordered state and the order parameter is zero, i.e., $r=0$. In this case, the function $A(s,q)$ given by Eq.~(\ref{As1}) is independent on $q$ and $A(s) \equiv A(s,q)$.  Equation (\ref{d17}) for the poles takes a simple form,
\begin{equation}
\frac{\langle q^2 \rangle A(s)}{\langle q\rangle}=\frac{1}{K}.
\label{d17-b}
\end{equation}

\paragraph*{Lorentz distribution of natural frequencies.} First let us find the relaxation rate in the case of the Lorentz distribution of natural frequencies with the zero mean value,
%%%
\begin{equation}
g(\omega) = \frac{\Delta}{\pi (\omega^2 + \Delta^2)}.
\label{Lorenzian}
\end{equation}
%%%
Throughout this paper, we will use $\Delta$ as a frequency unit and put $\Delta=1$.
Using this distribution function, we find
\begin{equation}
A(s) = \frac{1}{2}\int_{-\infty}^{\infty} d\omega \frac{g(\omega)}{s +  i \omega}= \frac{1}{2(1+s)},
\label{As2}
\end{equation}
that is the Eq.~(\ref{d17-b}) takes a form,
\begin{equation}
\frac{\langle q^2 \rangle}{2\langle q\rangle (1+s)} = \frac{1}{K}.
\label{perS}
\end{equation}
%%%
This equation has a single solution,
\begin{equation}
s_0 = - \frac{K_c-K}{K_c},
\label{d22}
\end{equation}
where the critical coupling $K_c=2\langle q\rangle/\langle q^2 \rangle$ [see Eq.~(\ref{d10})]. The inverse Laplace transform of Eq.~(\ref{d15}) yields
\begin{equation}
\delta z(t) \sim e^{-t/\tau_r}
\label{d23}
\end{equation}
where the relaxation rate $\tau_{r}^{-1}=-s_0$, i.e.,
\begin{equation}
\tau_{r}^{-1}=\frac{K_c-K}{K_c}.
\label{rr1}
\end{equation}
Note that this result is exact for a complete graph
%. It agrees with the one reported in
\cite{strogatz1991stability,Daido_1990,Ott2008}. The relaxation rate  $\tau_{r}^{-1}$ tends to zero when $K \rightarrow K_c$. This is critical slowing down, meaning that the relaxation time $\tau_{r}$ goes to infinity at the critical point.
%Note that the network structure determines only $K_c$.
The equation (\ref{rr1}) is valid for any uncorrelated complex network with a finite second moment $\langle q^2 \rangle$ when $K_c >0$, i.e., at the degree exponent  $\gamma >3$.
%when $K_c >0$.

\begin{figure}
\includegraphics[width=0.4\textwidth]{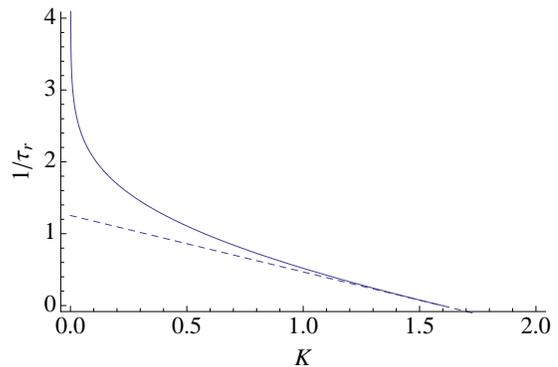}
\caption{Relaxation rate $1/\tau_r$ versus coupling $K$ in the Kuramoto model on a complete graph in the region $K < K_c$. The solid curve represents a numerical solution of Eq.~(\ref{d17-b}) in the case of a Gaussian distribution of the natural frequencies with the mean frequency $\langle \omega \rangle =0$ and the variance $\sigma=1$. The dashed line represents the linear approximation Eq.~(\ref{d29}).} \label{fig:gauss}
\end{figure}

%\paragraph*{Gaussian distribution of natural frequencies.}
%Thanks to the Central Limit theorem, Gaussian probability distribution, $g(\omega)=\exp(-\omega^2/2
%)/\sqrt{2\pi}$, is very widespread in nature, and deserves to be considered in the Kuramoto model as well. %
\paragraph*{Symmetric distributions of natural frequencies.}
In a general case of a symmetric frequency distribution $g(\omega)$, one can write the function $A(s)$ as  the Laplace transform,
\begin{equation}
A(s) = \int_0^\infty \hat{g}(y)e^{-sy} dy,
\label{d21}
\end{equation}
where $\hat{g}(y)$ is the Fourier transform of $g(\omega)$.
An analytical result for a solution $|s_0|\ll 1$ of Eq.~(\ref{d17-b}) can be obtained using the linearization
\begin{equation}
A(s) \approx \int_0^\infty \hat{g}(y) d y - s \int_0^\infty y\hat{ g}(y) d y
\label{d24}
\end{equation}
and assuming that these two integrals exist.
Introducing the coefficient $C=\int_0^\infty y\hat{ g}(y) d y$, the pole $s_0$ is given by
\begin{equation}
\frac{2}{K}\frac{\langle q \rangle}{\langle q^2 \rangle}=\pi g(0)-Cs_0.
\label{d26}
\end{equation}
Since the critical coupling equals $K_c=2 \langle q \rangle/(\pi g(0) \langle q^2 \rangle)$, we find the relaxation rate $\tau_{r}^{-1}$,
\begin{equation}
\tau_{r}^{-1} = -s_0  = \frac{\pi g(0)}{C} \frac{K_c-K}{K}.
\label{d28}
\end{equation}
Thanks to the Central Limit theorem, Gaussian probability distribution, $g(\omega)=\exp(-\omega^2/2)/\sqrt{2\pi}$, is very widespread in nature, and deserves to be considered in the Kuramoto model as well. In the Gaussian case, $C=1$ and
\begin{equation}
\tau_{r}^{-1} \approx \sqrt{\frac{\pi}{2}}\frac{K_c-K}{K_c}.
\label{d29}
\end{equation}
This critical behavior of $\tau_{r}^{-1}$ agrees with a numerical solution of Eq.~(\ref{d17-b}) displayed in Fig.~\ref{fig:gauss}.
%%%%%%%%%%%%%%%%%%%%%%%%%%%%%%%%%%%%%%%%%%%%  figure
%%%%%%%%%%%%%%%%%%%%%%%%%%%%%%%%%%%%%%%%%%%%%%

\subsection{Relaxation rate above the critical coupling, $K \geq K_c$}
\label{relaxation above}

Above the critical point, $K>K_c$, the order parameter is nonzero, $r\neq 0$, and we must substitute the function $\alpha_{0}(\omega,q)$ from Eq.~(\ref{d9}) into Eq.~(\ref{As1}).
For the Lorentz frequency distribution Eq.~(\ref{Lorenzian}), bulky but simple calculations give a simple result,
\begin{equation}
A(s,q)=\frac{\sqrt{1+(Kqr)^2}-1}{(Kqr)^2 [\sqrt{1+(Kqr)^2}+s]}.
\label{d30}
\end{equation}
Therefore, the equation (\ref{d17}) for poles takes a form,
\begin{equation}
\frac{1}{\langle q \rangle K^2 r^2}\int dq  P(q) \frac{\sqrt{1+(Kqr)^2}-1}{\sqrt{1+(Kqr)^2}+s} = \frac{1}{K}
\label{d17b}
\end{equation}

First, we consider the complete graph. Using $P(q)=\delta(q-N+1)$ and replacing $K (N-1) \rightarrow K$, we can write Eq.~(\ref{d17b}) in a form,
\begin{equation}
\frac{\sqrt{1+(Kr)^2}-1}{(Kr)^2 [\sqrt{1+(Kr)^2}+s]}=\frac{1}{K}.
\label{d31}
\end{equation}
Using the implicit solution Eq.~(\ref{op4}) for $r$,
%and $K_c=2$,
we find a solution $s_0$ and
the relaxation rate,
\begin{equation}
\tau_{r}^{-1} = -s_0 = K-2 =2 \varepsilon
\label{d32}
\end{equation}
where $\varepsilon = (K-K_c)/K_c$ and $K_c=2$. This simple result is exact at any $K\geq K_c$ for the complete graph. According to Eqs.~(\ref{rr1}) and (\ref{d32}), the relaxation rate $\tau_{r}^{-1}$ has the same critical index both below and above $K_c$, as one could expect from the mean-field theory. This result is in contrast to the work \cite{Daido_1990} where an asymmetric critical behavior was found.
The only difference between the critical behavior above and below $K_c$
%that we found
is that in the ordered phase the slope of the relaxation rate $\tau_{r}^{-1}$ is two times larger than one in the disordered phase [compare between Eq.~(\ref{rr1}) and Eq.~(\ref{d32})].
%This result is in complete agreement with the Ising model on a complete graph.

Now we find the relaxation rate in scale-free networks.
Assuming $|s| \ll 1$ in Eq.~(\ref{d17b}), we easily find
\begin{equation}
s_0 \simeq -\frac{T_1(r)}{T_2(r)}
\label{d33}
\end{equation}
where the integrals $T_1(r)$ and $T_2(r)$ are
\begin{eqnarray}
T_1(r) &=& \int_{q_0}^{\infty} d q P(q) \frac{(\sqrt{1+(Kqr)^2}-1)^2}{\sqrt{1+(Kqr)^2}},
\nonumber \\
T_2(r) &=& \int_{q_0}^{\infty} d q P(q) \frac{(\sqrt{1+(Kqr)^2}-1)}{1+(Kqr)^2}.
\label{d34a}
\end{eqnarray}
Here $q_0$ is the minimum degree in the network.

First we consider networks with a finite 4-th moment $\langle q^4 \rangle$ (i.e., $\gamma >5$ for scale-free degree distributions). In order to find $T_1(r)$ and $T_2(r)$, we use an expansion over $r \ll 1$. In the leading order in $r$, it gives
\begin{equation}
\tau_{r}^{-1}= - s_0 \simeq \frac{\langle q^4 \rangle K^2r^2}{2\langle q^2 \rangle}= 2 \varepsilon,
\label{d34b}
\end{equation}
where we used Eq.~(\ref{op2}). Therefore, at $\gamma >5$ the relaxation rate near the critical coupling $K_c$ is the same as for the complete graph (compare to Eq.~(\ref{d34b}) with Eq.~(\ref{d32})).

In the case $3 < \gamma < 5$, we find
\begin{eqnarray}
T_1(r) &\simeq & C (Kr)^{\gamma -1} a,
\nonumber \\
T_2(r) &\simeq & \frac{1}{2}K^2 r^2 \langle q^2 \rangle ,
\label{d35a}
\end{eqnarray}
where
\begin{equation}
a=\int_{0}^{\infty} dy\frac{y^{4-\gamma}}{\sqrt{1+y^2}(1+\sqrt{1+y^2})^2}
\label{d34d}.
\end{equation}
Then Eqs.~(\ref{op2}) and (\ref{d33}) give us the relaxation rate,
\begin{equation}
\tau_{r}^{-1} \simeq A \varepsilon,
\label{d34e}
\end{equation}
which is proportional to $\varepsilon$, similar to Eq.~(\ref{d34b}),  but with a coefficient $A$ different from 2.

Based on the results obtained above, we conclude that, despite network heterogeneity, the relaxation rate has the critical behavior, $\tau_{r}^{-1} \propto |(K-K_c)/K_c| $,  at any $\gamma >3$, both below and above the critical coupling $K_c$ in contrast to the critical dependence of the order parameter that has a non-mean field critical exponent at $3 < \gamma < 5$. Our results are summarized in the Table \ref{table1}.
Note that, in scale-free network with  $\gamma \leq 3$, the critical coupling is zero, i.e., $K_c =0$, \cite{Lee2005,Ichinomiya2004,Dorogovtsev2008}. In this case, the phase oscillators are synchronized at any non-zero coupling $K$ and the relaxation rate is finite.

\section{Kuramoto model in an external field and the susceptibility}
\label{susceptibility}

In the case of continuous phase transitions, critical slowing down and the divergence of a zero-field susceptibility are interrelated phenomena. The mean-field theory predicts that near the critical point the susceptibility is inversely proportional to the relaxation rate $\tau_{r}^{-1}$ found in the Sec.~\ref{rexation}.
In this section we show that this relationship is also valid for the Kuramoto model.

Let us introduce an external field into the Kuramoto model. Within the model, a phase oscillator can be described by two dimensional unit vector $\overrightarrow{n}(\theta)=(\cos \theta,\sin \theta)$. We introduce a field $\overrightarrow{h}(\phi)=h(\cos \phi,\sin \phi)$ characterized by an angle $\phi$ and a magnitude $h$. In a general case, $\phi$ and $h$ can be time dependent.
%and be different for different oscillators.
The energy of interaction between oscillator $i$ and a local field $\overrightarrow{h}_i$ is $E=-\overrightarrow{h}_i \overrightarrow{n}_i$. The field acts with a force $- \partial E/ \partial \theta_i= h_i \sin (\phi_i - \theta_i)$ on the oscillator. Substituting this force into the right hand side of Eq.~(\ref{a1}),  we obtain  equations describing dynamics of the Kuramoto model in the presence of local fields $\overrightarrow{h}_i$ \cite{Shinomoto01051986}: \begin{equation}
\frac{d \theta_i}{dt} = \omega_i + K\sum_{j=1}^N a_{ij}\sin(\theta_j-\theta_i) + h_i \sin (\phi_i - \theta_i).
\label{d35b}
\end{equation}
The local field $\overrightarrow{h}_i$ forces the
%the phases $\theta_i$ of the
oscillator $i$ to be parallel to the field. This model was studied in \cite{Shinomoto01051986,Sakaguchi1988,Strogatz1989,Arenas1994,Coolen2003,Acebron2005}.

Let us consider a uniform field, i.e., $h_i=h$ and $\phi_i =\phi$.
In general case, the field$\overrightarrow{h}$ can rotate with an angular velocity $\Omega_h$, i.e., $\phi = \Omega_h t +\phi_0$.  In this paper, we only study the resonance case when $\Omega_h$ equals the angular group velocity $\Omega$, i.e., $\Omega_h=\Omega$.
Under this choice in the rotating frame, the field is constant and determines
the phase $\psi$ of the order parameter $z$,
%in the rotating frame is equal to
i.e., $\psi=\phi_0$. For simplicity we chose $\phi_0 =0 $ and, therefore, $z=r$ in the steady state. We define the susceptibility as follows:
\begin{equation}
\chi(h)\equiv \frac{d r(h)}{dh},
\label{d35d}
\end{equation}
where the order parameter $r(h)$ must be found taking into account the field $h$. This definition is similar to the longitudinal susceptibility in the classical XY model. Another definition of the susceptibility was used in \cite{Coolen2003}.

In order to obtain $r(h)$, it is necessary to find the function $\alpha_0(\omega,q,h)$ in the steady state. One easily sees that $\alpha_0(\omega,q,h)$ is determined by Eq.~(\ref{d9}) if we replace $Kqr$ to $Kqr + h$. Then $r(h)$ is given by Eq.~(\ref{eq:scrq}) with $Kqr \rightarrow Kqr + h$ [see Eq.~(\ref{op-h1}) in the Appendix \ref{app:orderparameter-h}]. Note that $r(h)$ is nonzero at any coupling $K >0$  if $h\neq 0$.
Differentiating Eq.~(\ref{op-h1}) with respect to $h$ gives
\begin{equation}
\chi(h)=\frac{X_1(h)}{1-K X_2(h)},
\label{susc1}
\end{equation}
where
\begin{eqnarray}
X_1(h) &{=} & \frac{2}{\langle q \rangle } \int \frac{d q P(q) q}{(Kqr{+}h)^3 }  \int_{0}^{Kqr{+}h} \!\!\!\! d\omega \frac{\omega^2 g(\omega)}{\sqrt{1{+}[\omega /(Kqr{+}h)]^2}},
\nonumber \\
X_2(h) &{=} & \frac{2}{\langle q \rangle } \int \frac{d q P(q) q^2}{(Kqr{+}h)^3 }  \int_{0}^{Kqr{+}h} \!\!\!\! d\omega \frac{\omega^2 g(\omega)}{\sqrt{1{+}[\omega /(Kqr{+}h)]^2}}.
\nonumber \\
\label{susc2}
\end{eqnarray}
This equation determines the susceptibility as a function of $K$ and $h$.
In the case of the complete graph and the Lorentz frequency distribution,
the susceptibility equals
\begin{equation}
\chi(h)=\frac{1}{3Kr^2+2hr-K+2},
\label{susc-cg2}
\end{equation}
where $r$ is a solution of the cubic equation (\ref{op-h3}) in Appendix \ref{app:orderparameter-h}. This result is exact at any $K\geq 0$  and $h \geq 0$.

%In complex networks,
At $K< K_c$ and $h=0$, the order parameter is zero, $r=0$, and Eqs. (\ref{susc2}) give that the zero-field susceptibility equals
\begin{equation}
\chi(0)=\frac{1}{2 (1 - K/K_c)},
\label{susc2b}
\end{equation}
where $K_c$ is given by Eq.~(\ref{d10}). This result is exact at any $K \leq K_c$ in the complete graph.

%%%%%%%%%%%%%%%%%%%%%  table   %%%%%%%%%%%%%%%%%%%%%%%%%%%%%%%%%%
%%%%%%%%%%%%%%%%%%%%%%%%%%%%%%%%%%%%%%%%%%%%%%%%%%%%%%%%%%%%%%%%%%
\begin{table}[t]
%%[tbp]
\caption{
Critical behavior of the Kuramoto model on complex networks with degree distribution $P(q) \propto q^{-\gamma}$ for the order parameter $r$, the relaxation rate $\tau_{r}^{-1}$, and the susceptibility $\chi$ as functions of the coupling $K$, $|1-K/K_c| \ll 1$, at zero field $h=0$, and versus a magnetic field $h$ at the critical coupling $K=K_c$.  The width of the Lorentz distribution of natural frequencies is 1.}
%%%%%%%%%
%%%%%%%%
\begin{tabular}{|c|c|c|c|}
  \hline
  $\text{Network }$ & $r$ & $\tau_{r}^{-1} $ & $\chi$ \\
  \hline
 \text{$\gamma >5$} & & &
 \\
 $K<K_c$ & 0 & $ 1{-}\frac{K}{K_c} $ & $\frac{1}{2} \Bigl(1 {-} \frac{K}{K_c} \Bigr)^{-1}$
 \\
 $K > K_c$ & $\Bigl(\frac{K}{K_c} {-} 1\Bigr)^{1/2}$ & $ 2 \Bigl(\frac{K}{K_c} {-} 1\Bigr)$ & $\frac{1}{4} \Bigl(\frac{K}{K_c} {-} 1 \Bigr)^{-1}$
 \\
  $K=K_c$ & $\propto h^{1/3}$ &  $\propto h^{2/3}$ & $\propto h^{-2/3}$
  \\
  \hline
  \text{$3 < \gamma  \leq 5$}  & & &
  \\
  $K<K_c$ & 0 & $ 1{-}\frac{K}{K_c} $ & $\frac{1}{2} \Bigl(\frac{K}{K_c} {-} 1 \Bigr)^{-1}$
  \\
 $K > K_c$ & $\Bigl(\frac{K}{K_c} {-} 1\Bigr)^{1/(\gamma-3)}$ &  $\propto \Bigl(\frac{K}{K_c} {-} 1\Bigr)$ & $\frac{1}{2(\gamma -3)} \Bigl(\frac{K}{K_c} {-} 1 \Bigr)^{-1}$
 \\
 $K=K_c$  & $ \propto h^{1/(\gamma - 2)}$ &  $\propto h^{(\gamma - 3)/(\gamma - 2)}$ & $\propto h^{-(\gamma - 3)/(\gamma - 2)}$
 \\ \hline
\end{tabular}
\label{table1}
\end{table}

Above $K_c$ when $K-K_c\ll K_c$, we find
\begin{equation}
\chi(0) = \begin{cases} \frac{1}{4 (K/K_c -1)}, &  \gamma >5 \\
\frac{1}{2 (\gamma-3) (K/K_c -1)}, & 3<\gamma \leq 5.
\end{cases}
\label{susc3}
\end{equation}
Therefore, the network heterogeneity does not affect the critical exponent of the zero-field susceptibility if a network has a finite second moment of degree distribution. This critical behavior of susceptibility is a general property of mean-field models \cite{Stanley_book}. The heterogeneity affects only the numerical coefficient in Eq.~(\ref{susc3}). At $3<\gamma \leq 5$, we have the coefficient $1/[2 (\gamma-3)]$ instead of 1/4 at $\gamma >5$.

Let us find a field dependence of the susceptibility $\chi(h)$ at the critical point $K=K_c$. Using Eq.~(\ref{op-h5}) in the Appendix \ref{app:orderparameter-h} which determines  $r(h)$ at the critical coupling $K=K_c$,  we obtain
\begin{equation}
\chi(h) \propto \begin{cases} h^{-2/3} , &  \gamma >5 \\
h^{-(\gamma-3)/(\gamma-2)} , & 3<\gamma \leq 5.
\end{cases}
\label{susc-h}
\end{equation}
One can see that network heterogeneity changes the critical exponent of the susceptibility $\chi(h)$ if the fourth moment of the degree distribution diverges. This result is in contrast to the fact that the critical behavior of  the zero-field susceptibility $\chi(0)$ is not affected by the heterogeneity [see Eq.~(\ref{susc3})].

Our results are summarized in the Table \ref{table1}. One can see that the Kuramoto model on uncorrelated random complex networks has the same critical exponents as the Ising model, i.e., it belongs to the same class universality as the Ising model \cite{DGM_2002,Dorogovtsev2008}.

\section{Simulations of the Kuramoto model on complex networks}
\label{simulations}

\begin{figure}
\includegraphics[width=0.3\textwidth]{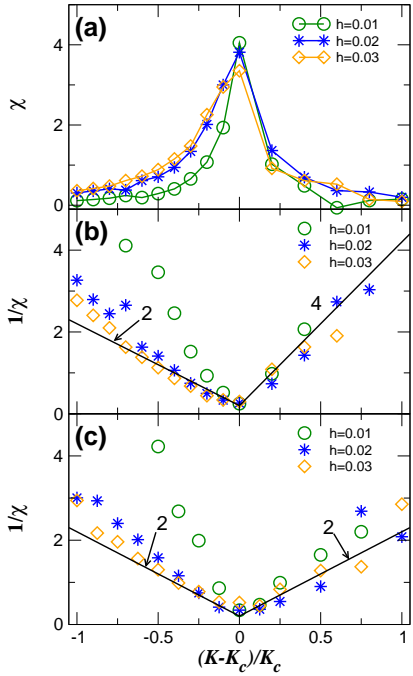}
\caption{(Color online) Susceptibility $\chi$, Eq.~(\ref{d35d}), versus the coupling $K$ near the critical point $K_c$ in the Kuramoto model on scale-free networks. Results of numerical simulations: (a) $\chi$ in a scale-free network with the degree exponent $\gamma=6$  in the fields $h=0.01$ (circles), $h=0.02$ (stars), and $h=0.03$ (diamonds). Panels (b) and (c) show $1/\chi$ at  $\gamma=6$ and 4, respectively. Solid lines are lines $A(K_c-K)/K_c$ and $B (K-K_c)/K_c$ with the slopes $A=2$ and $B=4$ in the panel (b) and $A=2$ and $B=2$ in the panel (c). The mean degree of the networks is $\langle q \rangle=20$. Each  point on the plots is obtained by averaging over $10^4$ time steps in the iterative procedure and 100 network realizations.
\label{fig-susceptibility}}
\end{figure}

In order to test our approach based on annealed network approximation,
we performed numerical simulations of the Kuramoto model on scale-free complex networks. The natural frequencies were distributed according to the Lorentz distribution, Eq.~(\ref{Lorenzian}), with $\Delta=1$.
We built scale-free complex networks with $N=10^4$ nodes by use of the static model \cite{GKK_2001,LGKK_2004} with the degree exponents $\gamma=4$ and 6 and the mean degree $\langle q \rangle =20$. Then we set the Kuramoto model on the networks and studied its dynamics by solving the set of rate equations (\ref{d35b}) in a uniform field $\overrightarrow{h}=(h,0)$. We used an iterative method with the time step $\Delta t =0.001$.
%, i.e., $\phi_i=\phi=0$.
In our simulations we measured the order parameter $r(h)$ given by Eq.~(\ref{op-o}), averaging $\sqrt{z(t)z^{*}(t)}$ over a large observation time.
%when the system reaches a steady state.
Then we found the susceptibility $\chi (h)=[r(h)-r(h-\Delta h)]/\Delta h$, Eq.~(\ref{d35d}). Furthermore, the physical parameters were averaged over 100 network realizations.
Results of our simulations  are displayed in Fig.~\ref{fig-susceptibility} for networks with finite and divergent fourth moment of the degree distribution, $\gamma=6$ and $\gamma=4$, respectively. Figure \ref{fig-susceptibility}(a) shows that the susceptibility  $\chi (h)$ has a sharp peak at $K=K_c \approx 0.1$ and this peak is growing when the field $h$ decreases. In Figs. \ref{fig-susceptibility} (b) and \ref{fig-susceptibility} (c) one can see that the reciprocal of the susceptibility  $1/\chi (h)$ is proportional to $| K-K_c|/K_c$ with the slopes that agree with our results in Table~\ref{table1}.

\section{Pair correlation functions and susceptibility}
\label{pair correlations}

In the classical XY model, if we apply a magnetic field $h_x$ along the $x$-axis, the longitudinal susceptibility $dM_x/dh_x$ is
\begin{eqnarray}
\frac{dM_x}{dh_x} &{=}& \langle M_{x}^2 \rangle - \langle M_x \rangle^2 ,
\nonumber \\
&{=}& \frac{1}{N}\sum_{i,j}[\langle \cos\theta_i \cos\theta_j \rangle_s {-} \langle \cos\theta_i \rangle_s \langle \cos\theta_j \rangle_s],
\label{susc-XY}
\end{eqnarray}
where $M_x=N^{-1}\sum_{i} \cos\theta_i $ is the $x$-component of the magnetization $\overrightarrow{M}=(M_x, M_y)$. Here $\langle \dots \rangle_s$ stands for the standard average over the statistical ensemble. This equation relates the susceptibility to the pair correlation function.

\begin{figure}
\includegraphics[width=0.4\textwidth]{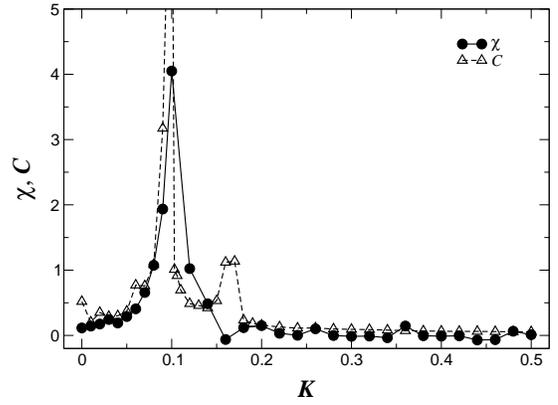}
\caption{(Color online) Susceptibility $\chi$, Eq.~(\ref{d35d}), and correlation function $C$, Eq.~(\ref{corr-f1}), versus coupling $K$ in the Kuramoto model on a scale-free network with the degree exponent $\gamma=6$ in the field $h=0.01$. In simulations, the mean degree of the network is $\langle q \rangle=20$.
%Each point in the plots is obtained by  averaging  over 100 network realizations.
\label{fig-corr-suscept}}
\end{figure}

In the case of the Kuramoto model, we suggest that a long time average,
\begin{equation}
\langle A(t) \rangle_t \equiv  \frac{1}{T}\int_{0}^{T} A(t) d t,
\label{Kuramoto-t-av}
\end{equation}
replaces $\langle A \rangle_s$ where the observation time $T \gg 1$ and the phases $\theta_i(t)$  of oscillators  are determined in the rotating frame related with an angular group  velocity $\Omega$. Introducing a field $\overrightarrow{h}=h(\cos\phi,\sin\phi)$ with the phase $\phi=0$, as it was discussed in Sec. \ref{susceptibility},
%alike Eq.~(\ref{susc-XY})
we define a pair correlation function in the Kuramoto model as follows:
\begin{eqnarray}
C(h) &{=}&  \frac{1}{N} \sum_{i,j}[\langle \cos\theta_i(t) \cos\theta_j(t) \rangle_t {-} \langle \cos\theta_i (t) \rangle_t \langle \cos\theta_j (t) \rangle_t]
\nonumber \\
 &{=}& N \{\langle \text{Re}^2 [z(t)]  \rangle_t - \langle \text{Re}  [z(t)] \rangle_t^{2} \}
\label{corr-f1},
\end{eqnarray}
where $\text{Re} [z(t)]=r(t)\cos[\psi(t)]$ is the real part of the complex parameter $z(t)$ in Eq.~(\ref{op-o}). The phase $\psi$ fluctuates around the field phase $\phi=0$ in the rotating frame.
We suggest that
the pair correlation function $C(h)$ is equal to the susceptibility $\chi(h)$, Eq.~(\ref{d35d}), i.e.,
\begin{equation}
C(h)=\chi(h).
\label{susc=corr_f}
\end{equation}
The equality of the susceptibility to a pair correlation function is a general property of statistical physics models.
Based on this suggestion and Table~\ref{table1}, we expect that $C(h=0) \propto |K-K_c|^{-1}$ both above and below $K_c$.

One can also introduce the pair correlation function in slightly different form,
\begin{eqnarray}
\widetilde{C}(h) &{=}&  N [\langle z(t) z^{*}(t) \rangle_t - \langle z(t) \rangle_t \langle z^{*}(t) \rangle_t],
\nonumber \\
&{=}& N [\langle r^{2}(t) \rangle_t  {-} \langle r(t)e^{i\psi(t)} \rangle_t \langle r(t)e^{-i\psi(t)} \rangle_t ].
\label{corr-f2}
\end{eqnarray}
Here, as well as above, $z(t)$ is defined in the rotating frame. Note that the function Eq.~(\ref{corr-f2}) is similar to the correlation function $N[\langle \overrightarrow{M}\overrightarrow{M} \rangle_s - \langle \overrightarrow{M} \rangle_s \langle \overrightarrow{M} \rangle_s]$ in the XY model. It is well known that the latter function has the same critical behavior as the function in Eq.~(\ref{susc-XY}).

The function Eq.~(\ref{corr-f2}) was introduced by Daido in the case of a complete graph \cite{Daido_1986}.
Numerical simulations in \cite{Daido_1986,Daido_1990,HPT_2007} revealed that $\widetilde{C}$ has a sharp maximum at the critical coupling $K_c$.
Using an analytical approach in \cite{Daido_1990}, Daido found that $\widetilde{C} \propto |K-K_c|^{-\gamma'}$ with the critical exponent $\gamma'=1$ when $K < K_c$
and $\gamma'=1/4$ when $K > K_c$. However, the origin of this asymmetry was not explained. Note that this result is in contrast to the mean field theory which predicts symmetric critical behavior with $\gamma'=1$. Note also that the analytical approach in \cite{Daido_1990} gives the relaxation rate $\tau_{r}^{-1} \propto (K - K_c)^{1/4}$ in contrast to our exact result $\tau_{r}^{-1} = K- K_c$ [see Eq.~(\ref{d32})].

\begin{figure}
\includegraphics[width=0.3\textwidth]{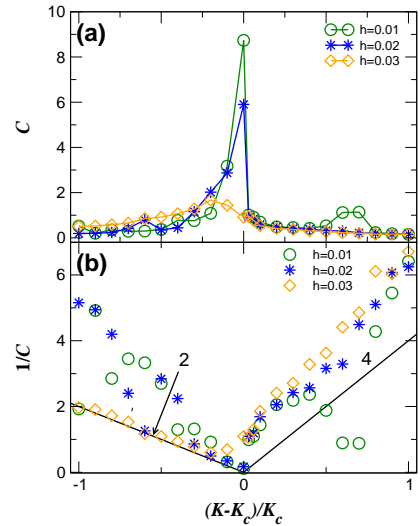}
\caption{(Color online) (a) Pair correlation function  $C$, Eq.~(\ref{corr-f1}), versus the coupling $K$ near the critical point $K_c$ in a scale-free network with the degree exponent $\gamma=6$ in the fields $h=0.01$ (circles), $h=0.02$ (stars), $h=0.03$ (diamonds). Panel (b) shows $1/C$. Solid lines are lines $A(K_c-K)/K_c$ and $B (K-K_c)/K_c$ with the slopes $A=2$ and $B=4$. In simulations, the mean degree of the network is $\langle q \rangle=20$.
\label{fig-corr-function}}
\end{figure}

In order to verify our suggestion Eq.~(\ref{susc=corr_f}), we solved numerically equations (\ref{d35b}) by use of the standard iterative method and calculated $C$ by use of  Eq.~(\ref{corr-f1}) in the case of a scale-free network with $N=10^4$ nodes and the degree exponent $\gamma=6$ in small fields $h$. Results of our simulations are represented in Figs. \ref{fig-corr-suscept} and \ref{fig-corr-function}. Each  point on the plots is obtained by integrating over $10^5$ time steps ($\Delta t= 0.001$) in Eq.~(\ref{Kuramoto-t-av}) (initial $5\times 10^5$ time steps in the iterative procedure were skipped in order to reach a steady state).
Figure~\ref{fig-corr-suscept} shows that $C(h)$ has a sharp peak at the critical point $K=K_c$ and, moreover, $C(h)\approx \chi(h)$. As one can see in Fig.~\ref{fig-corr-function}, near the critical coupling, $1/C$ is proportional to $|K-K_c|$ with approximately the same slope as $1/\chi$ in Table~\ref{table1} and Fig.~\ref{fig-susceptibility}.
We also calculated the function $\widetilde{C}$ and found that it has the same critical behavior as $C$, i.e., $\widetilde{C}(h=0) \propto |K-K_c|^{-1}$ with the same critical exponent both below and above $K_c$, but $\widetilde{C}(h=0)$ was approximately two times larger than $C$. We assume that this difference is due to phase fluctuations since the network size and the mean degree were not sufficiently large in our simulations. We expect that in the limit $N \rightarrow \infty$ and $\langle q \rangle \gg 1$, $C$ and $\widetilde{C}$ are equal to each other in the critical region at a non-zero field.

Thus, our simulations support the suggestion Eq.~(\ref{susc=corr_f}) that in the Kuramoto model the pair correlation function Eq.~(\ref{corr-f2}) is equal to the susceptibility Eq.~(\ref{d35d}). The divergence of $C(0)$ at the critical coupling signals emergence of infinite-range pair correlations at $K_c$. However, further  theoretical analysis and more detailed simulations of larger networks are necessary to prove this suggestion.

\section{Conclusion}

In this paper, we studied an impact of network heterogeneity on relaxation dynamics and
susceptibility of a system of coupled phase oscillators described by the Kuramoto model
on uncorrelated complex networks with scale-free degree distributions $P(q)\propto q^{-\gamma}$.
By use of the Ott-Antonsen method and the annealed network approach, we found analytically the critical behavior of the relaxation rate $\tau_{r}^{-1}$ and the susceptibility $\chi$ near (both above and below) the critical coupling $K_c$ of a second-order phase transition into a synchronized state. In particular we showed that near $K_c$ the susceptibility is inversely proportional to  $\tau_{r}^{-1}$ at all $\gamma >3$ in agreement of the mean-field theory.
We studied how the relaxation dynamics is influenced by the network heterogeneity. Our analysis showed that the critical behavior of the relaxation rate and the susceptibility follows the standard mean-field power laws, $\tau_{r}^{-1} \propto |K-K_c|$ and $\chi \propto 1/|K-K_c|$, if the second moment of the degree distribution is finite, i.e., at $\gamma >3$, in contrast to the critical behavior of the order parameter which is strongly influenced by the heterogeneity in networks with a divergent fourth moment of the degree distribution ($3 < \gamma \leq 5$). The divergence of the susceptibility manifests an infinite response of the Kuramoto model to an external field.
However, we found that a strong influence of the network structure can be observed in field dependence of the relaxation rate and the susceptibility when $3 < \gamma \leq 5$. At the critical coupling, these parameters are described by a power law with respect to an external field wherein the power law exponent depends on $\gamma$ when $3 < \gamma \leq 5$, in particular, $\chi(h)\propto h^{-(\gamma-3)/(\gamma-2)}$ in contrast to $\chi(h)\propto h^{-1/3}$ when $\gamma >5$.  Our results are summarized in Table~\ref{table1}. Furthermore,
we introduced a pair correlation function of phase oscillators and showed numerically that it has a sharp peak at the critical coupling,  signaling emergence of long-range correlations between oscillators.
We also carried out numerical simulations of the Kuramoto model on scale-free complex networks with finite and divergent forth moments of degree distributions.
Our simulations confirmed the critical behavior of the susceptibility and the pair correlation function. In particular, we demonstrated a strong enhancement of the pair correlation function and the susceptibility when the model approaches the critical point. This enhancement signals the emergence of long range correlations between oscillators.

\section{Acknowledgements}
This work was partially supported by FET IP Project MULTIPLEX 317532,
the PTDC projects SAU-NEU/ 103904/2008,  FIS/ 108476 /2008, MAT/ 114515 /2009, the project PEst-C / CTM / LA0025 / 2011, and the project `New Strategies Applied to Neuropathological Disorders,' cofunded by QREN and EU.
The work of M.~S.~S. was made possible by a scholarship supported by the European Commission within the framework of the \textsc{llp}-Erasmus Programme.

%\appendix
%\section{Steady states and the order parameter}
%\label{steadystates}

\appendix
\section{The order parameter versus the coupling $K$.}
\label{app:orderparameter-K}

In this appendix we develop a method that allows us to solve Eq.~(\ref{eq:scrq}) and find the order parameter $r$ as a function of the coupling $K$ in the case of scale-free degree distributions  $P(q)=Cq^{-\gamma}$, where $C$ is the normalization constant.
%It is this method that we use in the this paper in order to find the relaxation rate $\tau_r^{-1}$ and the susceptibility.
We consider a function,
\begin{equation}
\Phi(\rho) \equiv \frac{1}{\langle q \rangle}\int dq P(q) q^2 g(\rho q).
 %= \frac{K_c^{(0)}}{B+1}
\label{phi-ro}
\end{equation}
In the leading order in $\rho \ll 1$, this function has the following expansion with respect to $\rho$:
\begin{equation}
\Phi(\rho) = \begin{cases} \Phi_0 + \Phi_2 \rho^2 , &  \gamma >5 \\
\Phi_0 + \Phi_s \rho^{\gamma - 3}, & 3 < \gamma <5
\end{cases}
\label{phi-exp}
\end{equation}
where the coefficients are
\begin{eqnarray}
\Phi_0&=& g(0)\langle q^2 \rangle /\langle q \rangle,
\nonumber \\
\Phi_2 &=& g''(0) \langle q^4 \rangle /(2 \langle q \rangle),
\nonumber \\
\Phi_s &=& \frac{C}{\langle q \rangle (\gamma-3)(\gamma-4)}\int_{0}^{\infty}dx x^{4-\gamma} g''(x).
\label{coeff}
\end{eqnarray}
Here $g''(\omega)=d^2 g(\omega)/d\omega^2$ and $\langle q^n \rangle \equiv \sum_q P(q) q^n$ is the $n-$th moment of the degree distribution.

Using the Lorentz distribution Eq.~(\ref{Lorenzian}) for natural frequencies
%, $g(\omega)=\Delta / (\omega^2 + \Delta^2)$,
with $\Omega = \langle \omega \rangle =0$ and substituting the expansion (\ref{phi-exp}) into Eq.~(\ref{eq:scrq}), we find the critical coupling,
\begin{equation}
K_c=\frac{2}{\pi \Phi_0}=\frac{2\langle q\rangle}{\langle q^2 \rangle}
\label{d10}
\end{equation}
and the order parameter $r(K)$ as a function of $K$,
\begin{equation}
r = \begin{cases} \sqrt{\frac{\langle q \rangle^3}{\langle q \rangle^2 \langle q^4 \rangle}}\varepsilon^{1/2}, &  \gamma >5 \\
r_0 \varepsilon^{1/(\gamma-3)}, & 3<\gamma<5
\end{cases}
\label{op2}
\end{equation}
where $\varepsilon \equiv (K-K_c)/K_c$ and
\begin{equation}
r_0\equiv1/\Bigl(2K_{c}^{\gamma-2} |\Phi_s|\int_{0}^{1}dx x^{\gamma-3}\sqrt{1-x^2} \Bigr).
\label{op3}
\end{equation}
This result shows that the network heterogeneity strongly influences the critical exponent $\beta$ of the order parameter $r \propto \varepsilon^\beta$ in agreement with \cite{Lee2005,Ichinomiya2004,Dorogovtsev2008}. In networks with the degree exponent $3<\gamma<5$, when the fourth moment $\langle q^4 \rangle$ diverges,
the exponent $\beta$ differs from the mean-field value 1/2 that takes place at $\gamma >5$.
At $\gamma \leq 3$  the critical coupling $K_c=0$ and at any finite $K > 0$ the system is in the synchronized state.

Equations (\ref{eq:scrq})  and (\ref{eq:omq}) become exact for the complete graph. In this case, all oscillators have the same degree $q=N-1$ and $P(q)=\delta(q-N+1)$. Moreover, it is necessary to replace $Kq$ to $K$. For the Lorentz frequency distribution, the implicit solution of Eq.~(\ref{eq:scrq}) for the order parameter is
\begin{equation}
r=\sqrt{\frac{K-K_c}{K}},
\label{op4}
\end{equation}
where $K_c=2$.

Note that, in scale-free network with  $\gamma \leq 3$, the critical coupling is zero, $K_c =0$. and  the phase oscillators are synchronized, i.e., $r >0$, at any $K>0$ \cite{Lee2005,Ichinomiya2004,Dorogovtsev2008}.

\section{The order parameter versus an external field.}
\label{app:orderparameter-h}

Let us consider of the Kuramoto model in external field $h$. According to Sec. \ref{susceptibility}, in the rotating frame, the order parameter $r$  as a function of the coupling $K$ and the external field $h$ is determined by an equation,
%%%
\begin{equation}
r = \int \frac{d qP(q) q }{\langle q\rangle} \int_{-Kqr-h}^{Kqr + h} d\omega g(\omega)\sqrt{1{-}\left(\frac{\omega}{Kqr + h}\right)^2}.
\label{op-h1}
\end{equation}
Using the Lorentz distribution $g(\omega)$, Eq.~(\ref{Lorenzian}), we find that this equation takes a simple form,
\begin{equation}
r = \int \frac{d qP(q) q }{\langle q\rangle} \frac{[\sqrt{1+(Kqr + h)^2}-1]}{(Kqr + h)}.
\label{op-h2}
\end{equation}
In the case of the complete graph, i.e., when $P(q)=\delta(q-N+1)$ and $K(N-1) \rightarrow K$, Eq.~(\ref{op-h2}) is reduced to a cubic equation,
\begin{equation}
K r^3 +h r^2 - (K-2)r - h=0.
\label{op-h3}
\end{equation}
At $h=0$, a solution of the cubic equation is given by Eq. (\ref{op4}). At $K=K_c=2$ and $h \ll 1$ , we find
\begin{equation}
r \approx \Big(\frac{h}{K_c} \Bigr)^{1/3}.
\label{op-h4}
\end{equation}
%where we used $K_c=2$.
Interestingly, the magnetization $M$ of the  Ising model on a regular Bethe lattice has the same critical behavior $M \propto h^{1/3}$ \cite{Baxter_book}. Differentiating Eq.~(\ref{op-h3}) with respect to $h$ gives the explicit susceptibility for the complete graph,
\begin{equation}
\chi(h)=\frac{dr}{dh}=\frac{1}{3Kr^2+2hr-K+2}.
\label{susc-cg3}
\end{equation}

In the case of a scale-free degree distribution, $P((q)=Cq^{-\gamma}$, solving Eq.~(\ref{op-h2}), we find
\begin{equation}
r = \begin{cases} \Bigl(\frac{\langle q^2 \rangle ^2}{\langle q^4 \rangle} \Bigr)^{1/3} \Bigl(\frac{h}{\langle q \rangle K_c} \Bigr)^{1/3} , &  \gamma >5 \\
\frac{1}{q_0 K_c}\Bigl(\frac{q_0 K_c}{\alpha} \Bigr)^{1/(\gamma-2)} \Bigl(\frac{h}{q_0 K_c} \Bigr)^{1/(\gamma-2)} , & 3<\gamma<5
\end{cases}
\label{op-h5}
\end{equation}
where the coefficient $\alpha$ is
\begin{equation}
\alpha \equiv (\gamma-2)\int_{0}^{\infty}d x \frac{x^{\gamma-4}}{(1+\sqrt{1+x^2})^2},
\label{op-h6}
\end{equation}
and $q_0$ is the minimum degree. Thus, the equation (\ref{op-h5}) shows that the network heterogeneity strongly impacts the field dependence  of the order parameter $r(h)$ and the susceptibility $\chi=d r(h)/dh$ at the critical point $K=K_c$ if the network has a divergent fourth moment of the degree distribution. Instead of the standard mean-field critical exponent 1/3 when $\gamma >5$, we obtain $1/(\gamma{-}2)$ when $3< \gamma <5$. This result is an agreement with the generalized Landau theory of phase transitions in complex networks \cite{Dorogovtsev2008}.

\bibliography{bibliography}

\end{document}